\def\Kunits{{\rm keV \, cm^2}}

\documentclass{emulateapj}
\usepackage{epsfig}

\begin{document}

\slugcomment{\apj\ Letters, submitted 22 Apr 2008, accepted 23 May 2008}
\title{Conduction and the Star-Formation Threshold in Brightest Cluster Galaxies}
\author{G. M. Voit\altaffilmark{1},
        K. W. Cavagnolo\altaffilmark{1},
        M. Donahue\altaffilmark{1},
        D. A. Rafferty\altaffilmark{2},
        B. R. McNamara\altaffilmark{3,4,5}, \\
        P. E. J. Nulsen\altaffilmark{5}
         } 
\altaffiltext{1}{Department of Physics and Astronomy,
                 Michigan State University,
                 East Lansing, MI 48824, 
                 voit@pa.msu.edu, donahue@pa.msu.edu, cavagnolo@pa.msu.edu}
\altaffiltext{2}{Department of Astronomy and Astrophysics, 
                 Penn State University,
                 University Park, PA 16802, 
                 rafferty@astro.psu.edu}
\altaffiltext{3}{Department of Physics and Astronomy, 
                 University of Waterloo,
                 Waterloo, Ontario, Canada, N2L 3G1}
\altaffiltext{4}{Perimeter Institute for Theoretical Physics, 
                 31 Caroline St. N,
                 Waterloo, Ontario, Canada, N2L 2Y5}
\altaffiltext{5}{Harvard-Smithsonian Center for Astrophysics, 
                 60 Garden Street,
                 Cambridge, MA 01238}

\begin{abstract}
Current models of galaxy evolution suggest that feedback from active galactic nuclei is needed to explain the high-luminosity cutoff in the galaxy luminosity function.  Exactly how an AGN outflow couples with the ambient medium and suppresses star formation remains poorly understood.  However, we have recently uncovered an important clue to how that coupling might work.  Observations of H$\alpha$ emission and blue light from the universe's most luminous galaxies, which occupy the centers of galaxy clusters, show that star formation happens only if the minimum specific entropy of the intracluster gas is $\lesssim 30 \, {\rm keV \, cm^2}$.  Here we suggest that this threshold for star formation is set by the physics of electron thermal conduction, implying that conduction is critical for channeling AGN energy input toward incipient star-forming regions and limiting the progress of star formation.  
\end{abstract}

\section{Introduction}

\setcounter{footnote}{0}

What determines the upper limit to the luminosity of a galaxy?  This question has received a great deal of attention lately because the answer seems to involve feedback from a galaxy's nucleus.  Semi-analytic models of galaxy formation show that feedback of some kind is necessary to account for the observed characteristics of the galaxy luminosity function.  Models without feedback overpredict the number of both low-luminosity and high-luminosity galaxies.  Plausible amounts of feedback from supernovae bring predictions for low-luminosity galaxies into agreement with observations, but getting things right at the high-luminosity end is more difficult---the amount of supernova feedback needed to quench star formation in these systems is implausibly large \citep{Benson03}.  Numerical simulations of galaxy formation treat the physics of merging and cooling more realistically but still do not solve the problem of the high-luminosity cutoff.  In current simulations, late-time cooling of hot gas in the central galaxies of clusters produces so many stars at $z \sim 0$ that the galaxies have luminosities several times larger than the observed cutoff in the galaxy luminosity function and colors that are much too blue \citep[e.g.,][]{Saro06}.  

A closely related cooling problem has long bedeviled our understanding of galaxies and galaxy clusters, and X-ray observations strongly suggest that the solution lies with active galactic nuclei.  In roughly half of nearby clusters, the cooling time at $r \lesssim 100$~kpc is less than a Hubble time, implying that gas at the center of the cluster must cool and condense, if there is no mechanism to offset cooling.  However, the star-formation rates in these systems are at most 1\% to 10\% of the expected gas condensation rate, and X-ray spectroscopy shows that the the bulk of the central gas remains suspended at temperatures $\gtrsim 1/3$ of the virial temperature \citep[see][for a review]{pf06}.  High-resolution imaging with {\em Chandra} has revealed that many clusters and giant elliptical galaxies with short central cooling times also have cavities in their ICM that coincide with radio-emitting plasma outflows from the active nucleus in the central galaxy.  In many cases the AGN energy output inferred from these cavities is similar to the X-ray cooling rate of the central gas---a strong piece of circumstantial evidence in favor of AGN feedback as the mechanism that limits cooling and star formation in high-mass halos \citep{Birzan04,mn07}.  Not all clusters with short central cooling times have obvious cavities, but that finding can be understood if AGN heating is episodic.

The most recent versions of semi-analytic models have therefore incorporated schematic implementations of AGN feedback and find that ``radio-mode" feedback is a plausible explanation for both the observed cutoff in the galaxy luminosity function and the fact that the largest galaxies in the universe are red, not blue \citep{Croton06,Bower06}.  To match the observations, radio-mode feedback must occur preferentially in high-temperature systems and must provide enough feedback energy to suppress star formation without making many stars.  In this context, AGN feedback is most effective in systems experiencing hot-mode accretion because an extended hot galactic atmosphere is necessary to halt and thermalize the energy of the AGN outflow.

Here we consider the implications of two recent studies showing that star formation in brightest cluster galaxies (BCGs) is closely linked to the entropy\footnote{As is customary in the X-ray cluster field, we quantify entropy using the adiabatic constant $K = kT n_e^{-2/3}$ \citep[see][for more background]{Voit05}.} structure of the intracluster medium (ICM).  Section~2 presents evidence that star formation occurs only in BCGs whose central entropy is $\lesssim 30 \, {\rm keV \, cm^2}$, equivalent to a cooling time $\lesssim 5 \times 10^8$~years.  The star formation rates in these systems are generally much smaller than the gas cooling rates in the absence of feedback, and understanding why star formation still proceeds at a reduced rate is likely to tell us something about the coupling between AGN feedback and the ICM.  In section~3 we suggest that the critical central entropy threshold for star formation is determined by electron thermal conduction, because multiphase structure cannot persist in the ICM if the central entropy is much larger than the observed threshold.  If conduction is indeed responsible for governing the rate of star formation, then it is also likely to be a major conduit of thermal energy in the cluster core and a crucial part of the AGN feedback loop.  Section~4 summarizes the paper.

\begin{figure}[t]
\includegraphics[width=3.5in, trim = 1.2in 0.2in 1.6in 0.7in]{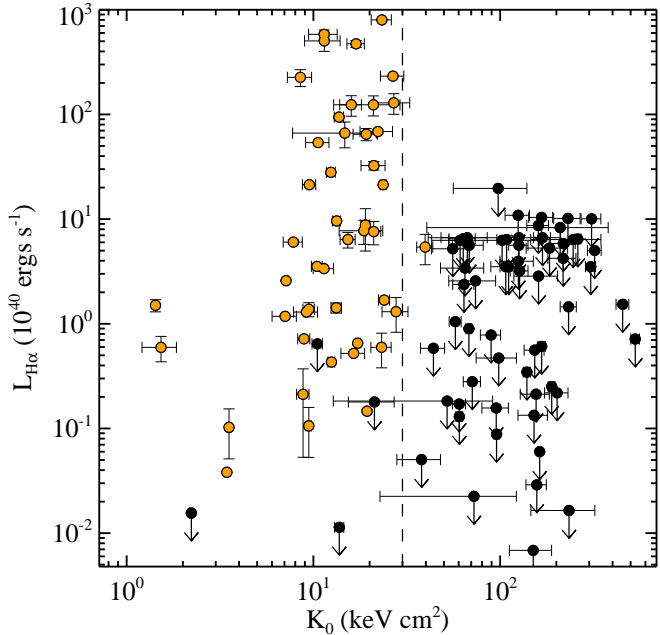} 
\caption{ \footnotesize
Dependence of H$\alpha$ emission in BCGs on central entropy from Cavagnolo et al. (2008a).  Central entropy $K_0$ has been measured by fitting {\em Chandra} entropy profiles with a radial power law plus a constant, and $L_{{\rm H}\alpha}$ values are a heterogeneous set taken from the literature.  Orange filled circles show clusters with detected H$\alpha$, and black filled circles give upper limits for clusters without detections (see Cavagnolo et al. 2008b for more details).  Notice that strong H$\alpha$ emission appears only when $K_0 \lesssim 30 \, {\rm keV \, cm^2}$.  
\label{Ha-K0}}
\end{figure}

\section{The Star-Formation Threshold in Brightest Cluster Galaxies}

Star formation in brightest cluster galaxies has long been known to correlate with the central cooling time of the host cluster.  \citet{hcw85} showed that H$\alpha$ nebulosity is present in a BCG only if the cluster's cooling time is less than a Hubble time, and much of the H$\alpha$ emission seen in these objects can be attributed to photoionization by hot stars \citep[e.g.,][]{jfn87,mo89,vd97}.  While it is possible that some of the H$\alpha$ comes from other sources like conductive interfaces \citep[e.g.,][]{Sparks04}, turbulent mixing layers \citep{bf90}, or cosmic rays \citep{r08,f08}, there is a clear connection between H$\alpha$ emission and the presence of extended excess blue light within the BCG \citep[e.g.,][]{Cardiel98}.  Furthermore, clusters with short central cooling times also often have substantial CO luminosities, indicating that they contain enough cool molecular gas to support the observed level of star-formation \citep{Edge01}.

Two recent studies have brought the star-formation threshold into sharper focus.  One is a {\it Chandra} archival survey of galaxy clusters by \citet{Cavagnolo08a} that measured the radial entropy profiles of 222 galaxy clusters, fitting them with the three-parameter expression
\begin{equation}
  K(r) = K_0 + K_{100} \left( \frac {r} {100 \, {\rm kpc}} \right)^\alpha \; \; .
 \label{eq-Kfit}
\end{equation}  
Here we will refer to $K_0$ as the central entropy, although in some cases it is not necessarily the minimum entropy level of the ICM but rather the entropy scale at which the radial profile departs from the best-fitting power law at large radii.  Figure~\ref{Ha-K0} shows that central H$\alpha$ emission has been detected only in clusters with $K_0 \lesssim 30 \, {\rm keV \, cm^2}$ (Cavagnolo et al. 2008b).   Emission-line luminosities in this plot have been taken from the literature and correspond to a variety of apertures, so the quantitative values of $L_{{\rm H}\alpha}$ are not very meaningful.  However, the entropy threshold for having detectable H$\alpha$ emission is quite distinct.  The presence of H$\alpha$ emission clearly indicates that the intracluster medium can have a multiphase structure below this entropy threshold, with a component at $\sim 10^4$~K in addition to the X-ray emitting component at $> 10^7$~K.  

\begin{figure}[t]
\includegraphics[width=3.1in, trim = -0.2in 0in 0.1in 0.0in]{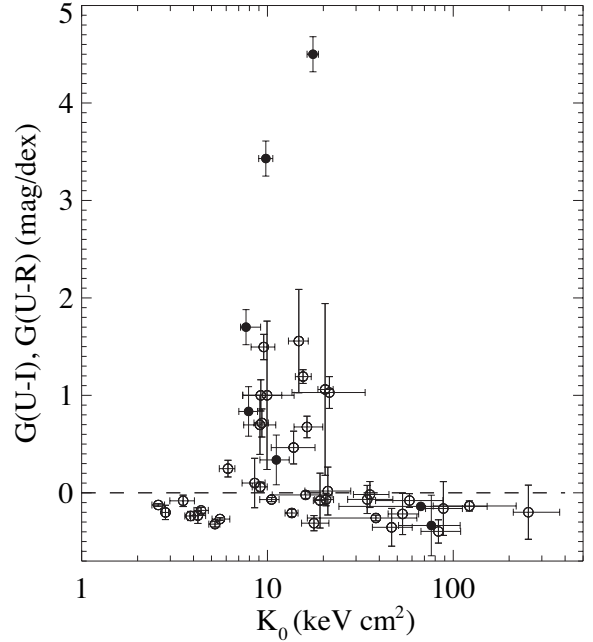} 
\caption{ \footnotesize
Dependence of BCG color gradient on central entropy from \citet{rmn08}.  Color gradients have been measured in $U-I$ and $U-R$, with open circles representing $G(U-I) = d(U-I)/d \log r$ and filled circles representing $G(U-R) = d(U-R)/d \log r$.  The dashed line indicates a zero color gradient. The only BCGs with unusually blue centers are those in clusters with central entropy $\lesssim 20 \, {\rm keV \, cm^2}$, indicating that star formation is present only in those BCGs below this central entropy threshold.
\label{Bluegrad-K0}}
\end{figure}

Observations of color gradients in BCGs have led to a similar conclusion.  In a sample of 46 clusters, \citet{rmn08} find that 34 of the 20 clusters with central entropy $< 30 \, {\rm keV \, cm^2}$ have BCGs with extended blue light in their cores consistent with ongoing star formation, while none of the clusters with central entropy $> 30 {\rm keV \, cm^2}$ have BCGs with colors bluer than expected from a passively evolving stellar population (see Figure~\ref{Bluegrad-K0}.)   Here the central entropy has been measured from a small X-ray aperture $\sim 2-3 \, {\rm arcsec}$ at the center of the cluster, and the color gradients have been measured in $U-I$ and $U-R$, with $G(U-I) = d(U-I)/d \log r$ and $G(U-R) = d(U-R)/d \log r$.    

The transition to H$\alpha$ emission and star formation is sharp in both cases, but it is not yet clear whether the threshold ought to be expressed in terms of central entropy or central cooling time.  Here we are presenting the results as a function of central entropy because we are going to interpret them in terms of the ICM entropy structure in the next section.  However, it is interesting that the central cooling times at which H$\alpha$ and blue light appear are substantially smaller than $10^{10}$~years.  Simply having gas that is able to cool within a Hubble time is not enough for star formation in a BCG.  Some other more restrictive condition must also be satisfied.

It is also interesting that in both figures there are clusters with central entropy $< 20 \, {\rm keV \, cm^2}$ and cooling time $< 3 \times 10^8 $~years that do not show evidence for star formation.   Low entropy and a correspondingly short cooling time are therefore not sufficient to guarantee star formation.  Rafferty et al. (2008) present evidence suggesting that star formation proceeds only if the AGN heating rate is less than the gas cooling rate and the central galaxy is close to the X-ray brightness peak.

\section{Conduction and the Conditions for a Multiphase ICM}

We suspect that the critical entropy threshold for multiphase gas and star formation in BCGs may result from electron thermal conduction.  Cool star-forming clouds should appear only in systems whose size is greater than a critical length scale, known as the {\em Field length}, below which thermal conduction smoothes out temperature inhomogeneities \citep{Field65,bm90}.  One can derive the Field length heuristically by considering thermal balance for a cool cloud of radius $r$ embedded in a medium of temperature $T$.  Electron thermal conduction sends energy into the cloud at a rate $\sim r^2 \kappa(T) \cdot T/r$, where $\kappa(T) = 6 \times 10^{-7} T^{5/2} f_c \, {\rm ergs \, s^{-1} K^{-1} cm^{-1}}$ is the Spitzer conduction coefficient and $f_c$ is a suppression factor depending on the magnetic field structure in the medium.  Radiative cooling can rid the cloud of energy at a rate $\sim r^3 n^2 \Lambda(T)$, where the cooling function $\Lambda(T) \propto T^{1/2}$ for $T > 2$~keV.  Cooling and conduction are therefore in approximate balance for systems with a radius of order the Field length,
\begin{equation}
  \lambda_{\rm F} \equiv \left[ \frac {T \kappa (T)} {n^2 \Lambda (T)} \right]^{1/2} 
    = 4  \, {\rm kpc} \;  \left[ \frac {K} {10 \, {\rm keV \, cm^2}} \right]^{3/2} f_c^{1/2} \; \; .
\end{equation}
Through a coincidence of scaling, the Field length is a function of entropy alone when free-free emission is the dominant cooling mechanism \citep{Donahue05}.

Figure~3 illustrates how this criterion translates into the entropy-radius plane.  The long-dashed lines give the loci of points for which $\lambda_{\rm F}(K)= r$, given suppression factors $f_c = 0.2$ and 1.  Magnetic suppression of conduction is a complicated and incompletely understood process, but most recent estimates have been in the range $f_c \approx 0.2$-$0.3$ \citep{Malyshkin01,nm01}. Below each line, gas within radius $r$ constitutes a subsystem with $r>\lambda_{\rm F}$. Conduction cannot stabilize that gas against cooling, allowing multiphase gas to persist and star formation to proceed.  Above each line is the region of stability, in which conduction leads to evaporation and homogeneity.

\begin{figure}[t]
\includegraphics[width=3.5in, trim = 0.8in 1.2in 0.7in 1.1in]{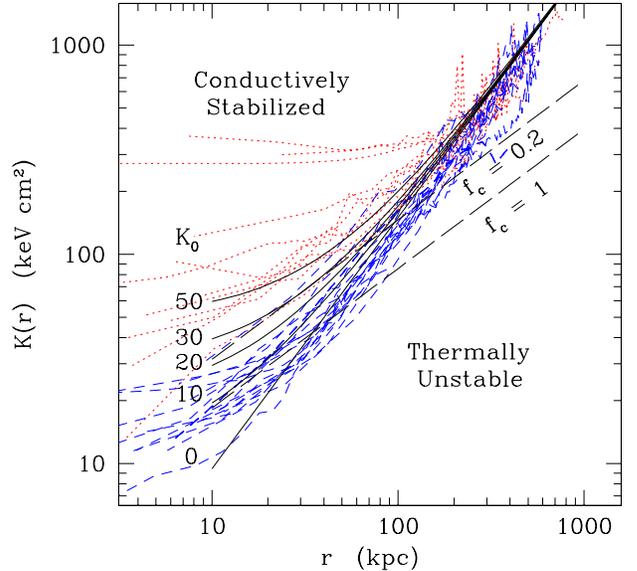} 
\caption{ \footnotesize
Regions of thermal stability in the ICM based on the Field-length criterion.   Long-dashed lines show where $\lambda_{\rm F}(K) = r$ for conduction suppression factors $f_c = 0.2$ and 1.  Above these lines, conduction should be more effective than radiative cooling, causing cooler structures of scale $<r$ to evaporate.  Below these lines, radiative cooling should be more effective than conduction, allowing thermal instability to proceed.  Solid lines show schematic entropy profiles of the form $K(r) = K_0 + (150 \, {\rm keV \, cm^2})(r/100 \, {\rm kpc})^{1.2}$, with $K_0 = 0$, 10, 20, 30, and 50, as labeled.   
Short-dashed blue lines give entropy profiles from (Cavagnolo et al. 2008a) for clusters in which Rafferty et al. (2008) find central star formation.  Dotted red lines give entropy profiles for clusters from Rafferty et al. (2008) without clear evidence for star formation and that also have no detectable H$\alpha$ emission (Cavagnolo et al. 2008b).
\label{kstable}}
\end{figure}

For comparison, the solid lines in Figure~3 show schematic intracluster entropy profiles motivated by the 
large {\em Chandra} entropy survey of Cavagnolo et al. (2008a).  Fitting the form in equation~(\ref{eq-Kfit}) to those profiles gives $K_{100} = 146.8 \pm 68.1 \, \Kunits$ and $\alpha = 1.22 \pm 0.26$ for clusters with $K_0 < 50 \, \Kunits$ and temperatures ranging from 2 keV to 12 keV.  The solid lines therefore represent typical intracluster entropy profiles with $K_{100} = 150 \, \Kunits$, $\alpha = 1.2$, and $K_0 = 0$, 10, 20, 30, and 50 keV cm$^2$.  Notice that only the schematic profiles with $K_0 < 30 \, {\rm keV \, cm^2}$, corresponding to clusters with central star formation and H$\alpha$ emission, dip below the threshold for conductive stabilization corresponding to $f_c = 0.2$.  Colored lines giving entropy profiles from (Cavagnolo et al. 2008a) reinforce this correspondence:  short-dashed blue lines show Rafferty et al. (2008) clusters with central star formation, and dotted red lines show Rafferty et al. (2008) clusters without clear evidence for star formation and without detectable H$\alpha$ emission (Cavagnolo et al. 2008b).

This simple analysis suggests that conduction is responsible for the strong dependence of H$\alpha$ emission and star formation on central entropy.  Even though the suppression factor $f_c$ is uncertain, a threshold must exist in the $K$-$r$ plane above which conduction heats and evaporates cooler gas clouds faster than radiative losses can cool them.  One therefore expects to see a transition from a multiphase ICM at low central entropy levels to a more homogeneous, isothermal ICM at higher entropy levels.  The H$\alpha$ observations are indicating that this transition is at approximately the expected location in the $K$-$r$ plane for thermal conduction with $f_c \approx 0.2$.  We note that the presence of such a threshold is independent of the origin of the line-emitting gas, which could come either from condensation of the ICM or from stripping of cool interstellar clouds from galaxies passing through the cluster core.  Stripped gas in an ambient medium above the critical $K(r)$ profile would quickly evaporate, while stripped gas in a medium below the profile could persist indefinitely.

\begin{figure}[t]
\includegraphics[width=3.5in, trim = 0.5in 0in 0in 0.2in]{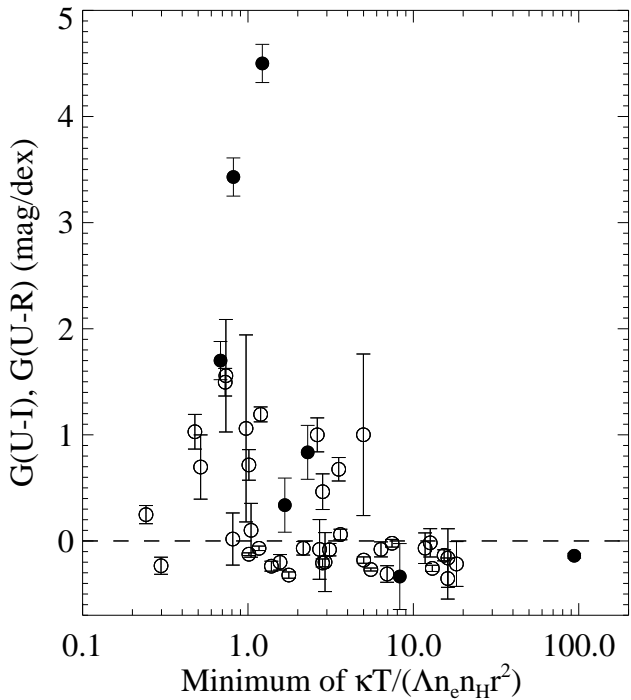} 
\caption{ \footnotesize
Dependence of BCG color gradient on conduction efficiency.  Open and filled circles have the same meanings as in Figure~\ref{Bluegrad-K0}.  Conduction efficiency is quantified by finding the minimum value of $\kappa T / (\Lambda(T) n_e n_{\rm H} r^2)$ for each cluster.  If this quantity is less than $1/f_c$, then conduction cannot offset radiative cooling.  The fact that blue gradients appear only in clusters with $\kappa T / (\Lambda(T) n_e n_{\rm H} r^2) \lesssim 5$ somewhere in the cluster suggests that thermal conduction with $f_c \approx 0.2$ determines whether a long-lasting multiphase medium can be present in the cluster core.
\label{Bluegrad-cond}}
\end{figure}

A case-by-case analysis of the effectiveness of conduction in the clusters from \citet{rmn08} agrees with the broader conclusions drawn from Figure~3.  For each cluster we have determined the minimum value of $\kappa T / [\Lambda(T) n_e n_{\rm H} r^2]$, where $\Lambda(T)$ represents the full cooling function and not just free-free emission, from {\em Chandra} observations. If $\kappa T / [\Lambda(T) n_e n_{\rm H} r^2] < f_c^{-1}$, then conduction cannot compensate for radiative cooling (see Figure 4). The fact that blue gradients are seen only in the BCGs of clusters with $\kappa T / [\Lambda(T) n_e n_{\rm H} r^2] \lesssim 5$ therefore provides further support for the idea that thermal conduction with $f_c \approx 0.2$ determines whether or not a BCG can form stars.

\section{Summary}

Two recent studies have shown that the star-forming properties of BCGs depend on the entropy structure of the host cluster's ICM.  Clusters with central entropy $\lesssim 30 \, \Kunits$, corresponding to a cooling time $\lesssim 5 \times 10^8$~years often have BCGs with H$\alpha$ emission (Cavagnolo et al. 2008b) and blue cores (Rafferty et al. 2008) indicative of ongoing star formation, while none of the clusters with high central entropy show the same features.  We suggest that this entropy threshold for star formation is set by the physics of thermal conduction.  Clusters typically have entropy profiles that approximately follow the law $K(r) = K_0 + (150 \, \Kunits)(r/ 100 \, {\rm kpc})^{1.2}$, with differing values of $K_0$  (Cavagnolo et al. 2008a).  If $K_0  > 30 \, \Kunits$, then conduction can carry heat energy toward the center of the cluster faster than the ICM can radiate it away, as long as the suppression factor $f_c \gtrsim 0.2$.  Therefore, we should not expect clusters with higher central entropy levels to have a persistent multiphase medium in their cores.  Conversely, conduction in lower-entropy clusters cannot fully compensate for radiative cooling, allowing star formation to proceed.  \citet{gor08} have recently presented a model explaining how AGN feedback can suppress star formation in these thermally unstable cluster cores.  The correspondence we find between the observed star-formation threshold and the predictions of conductive models indicates that conduction indeed is important in cluster cores and may be crucial to the AGN feedback mechanism thought to regulate star formation in these systems.  However, detailed hydrodynamical simulations including conduction will be necessary to determine more precisely how the phase structure of the ICM depends on a cluster's entropy profile.  

\vspace*{1.0em}

This work was partially supported at MSU by NASA grants NNG04GI89G and NNG05GD82G.  PEJN was supported by NASA grant NAS8-01130
 

\begin{thebibliography}{26}
\expandafter\ifx\csname natexlab\endcsname\relax\def\natexlab#1{#1}\fi

\bibitem[{{Begelman} \& {Fabian}(1990)}]{bf90}
{Begelman}, M.~C. \& {Fabian}, A.~C. 1990, \mnras, 244, 26P

\bibitem[{{Begelman} \& {McKee}(1990)}]{bm90}
{Begelman}, M.~C. \& {McKee}, C.~F. 1990, \apj, 358, 375

\bibitem[{{Benson} {et~al.}(2003){Benson}, {Bower}, {Frenk}, {Lacey}, {Baugh},
  \& {Cole}}]{Benson03}
{Benson}, A.~J., {Bower}, R.~G., {Frenk}, C.~S., {Lacey}, C.~G., {Baugh},
  C.~M., \& {Cole}, S. 2003, \apj, 599, 38

\bibitem[{{B{\^i}rzan} {et~al.}(2004){B{\^i}rzan}, {Rafferty}, {McNamara},
  {Wise}, \& {Nulsen}}]{Birzan04}
{B{\^i}rzan}, L., {Rafferty}, D.~A., {McNamara}, B.~R., {Wise}, M.~W., \&
  {Nulsen}, P.~E.~J. 2004, \apj, 607, 800

\bibitem[{{Bower} {et~al.}(2006){Bower}, {Benson}, {Malbon}, {Helly}, {Frenk},
  {Baugh}, {Cole}, \& {Lacey}}]{Bower06}
{Bower}, R.~G., {Benson}, A.~J., {Malbon}, R., {Helly}, J.~C., {Frenk}, C.~S.,
  {Baugh}, C.~M., {Cole}, S., \& {Lacey}, C.~G. 2006, \mnras, 370, 645

\bibitem[{{Cardiel} {et~al.}(1998){Cardiel}, {Gorgas}, \&
  {Aragon-Salamanca}}]{Cardiel98}
{Cardiel}, N., {Gorgas}, J., \& {Aragon-Salamanca}, A. 1998, \mnras, 298, 977

\bibitem[{{Cavagnolo} {et~al.}(2008){Cavagnolo}, {Donahue}, {Voit}, \&
  {Sun}}]{Cavagnolo08a}
{Cavagnolo}, K.~W., {Donahue}, M., {Voit}, G.~M., \& {Sun}, M. 2008a, in
  preparation

\bibitem[{{Cavagnolo} {et~al.}(2008){Cavagnolo}, {Donahue}, {Voit}, \&
  {Sun}}]{Cavagnolo08b}
{Cavagnolo}, K.~W., {Donahue}, M., {Voit}, G.~M., \& {Sun}, M. 2008b, ApJ Letters, submitted,
arXiv:0806.0382

\bibitem[{{Croton} {et~al.}(2006){Croton}, {Springel}, {White}, {De Lucia},
  {Frenk}, {Gao}, {Jenkins}, {Kauffmann}, {Navarro}, \& {Yoshida}}]{Croton06}
{Croton}, D.~J., {Springel}, V., {White}, S.~D.~M., {De Lucia}, G., {Frenk},
  C.~S., {Gao}, L., {Jenkins}, A., {Kauffmann}, G., {Navarro}, J.~F., \&
  {Yoshida}, N. 2006, \mnras, 365, 11

\bibitem[{{Donahue} {et~al.}(2005){Donahue}, {Voit}, {O'Dea}, {Baum}, \&
  {Sparks}}]{Donahue05}
{Donahue}, M., {Voit}, G.~M., {O'Dea}, C.~P., {Baum}, S.~A., \& {Sparks}, W.~B.
  2005, \apjl, 630, L13

\bibitem[{{Edge}(2001)}]{Edge01}
{Edge}, A.~C. 2001, \mnras, 328, 762

\bibitem[{{Ferland} {et~al.}(2008){Ferland}, {Fabian}, {Hatch}, {Johnstone},
  {Porter}, {van Hoof}, \& {Williams}}]{f08}
{Ferland}, G.~J., {Fabian}, A.~C., {Hatch}, N.~A., {Johnstone}, R.~M.,
  {Porter}, R.~L., {van Hoof}, P.~A.~M., \& {Williams}, R.~J.~R. 2008, \mnras,
  386, L72

\bibitem[{{Field}(1965)}]{Field65}
{Field}, G.~B. 1965, \apj, 142, 531

\bibitem[{{Guo} {et~al.}(2008){Guo}, {Oh}, \& {Ruszkowski}}]{gor08}
{Guo}, F., {Oh}, S.~P., \& {Ruszkowski}, M. 2008, arXiv:0804.3823

\bibitem[{{Hu} {et~al.}(1985){Hu}, {Cowie}, \& {Wang}}]{hcw85}
{Hu}, E.~M., {Cowie}, L.~L., \& {Wang}, Z. 1985, \apjs, 59, 447

\bibitem[{{Johnstone} {et~al.}(1987){Johnstone}, {Fabian}, \& {Nulsen}}]{jfn87}
{Johnstone}, R.~M., {Fabian}, A.~C., \& {Nulsen}, P.~E.~J. 1987, \mnras, 224,
  75

\bibitem[{{Malyshkin}(2001)}]{Malyshkin01}
{Malyshkin}, L. 2001, \apj, 554, 561

\bibitem[{{McNamara} \& {Nulsen}(2007)}]{mn07}
{McNamara}, B.~R. \& {Nulsen}, P.~E.~J. 2007, \araa, 45, 117

\bibitem[{{McNamara} \& {O'Connell}(1989)}]{mo89}
{McNamara}, B.~R. \& {O'Connell}, R.~W. 1989, \aj, 98, 2018

\bibitem[{{Narayan} \& {Medvedev}(2001)}]{nm01}
{Narayan}, R. \& {Medvedev}, M.~V. 2001, \apjl, 562, L129

\bibitem[{{Peterson} \& {Fabian}(2006)}]{pf06}
{Peterson}, J.~R. \& {Fabian}, A.~C. 2006, \physrep, 427, 1

\bibitem[{{Rafferty} {et~al.}(2008){Rafferty}, {McNamara}, \& {Nulsen}}]{rmn08}
{Rafferty}, D., {McNamara}, B., \& {Nulsen}, P. 2008, ArXiv e-prints, 802

\bibitem[{{Ruszkowski} {et~al.}(2008){Ruszkowski}, {En{\ss}lin}, {Br{\"u}ggen},
  {Begelman}, \& {Churazov}}]{r08}
{Ruszkowski}, M., {En{\ss}lin}, T.~A., {Br{\"u}ggen}, M., {Begelman}, M.~C., \&
  {Churazov}, E. 2008, \mnras, 383, 1359

\bibitem[{{Saro} {et~al.}(2006){Saro}, {Borgani}, {Tornatore}, {Dolag},
  {Murante}, {Biviano}, {Calura}, \& {Charlot}}]{Saro06}
{Saro}, A., {Borgani}, S., {Tornatore}, L., {Dolag}, K., {Murante}, G.,
  {Biviano}, A., {Calura}, F., \& {Charlot}, S. 2006, \mnras, 373, 397

\bibitem[{{Sparks} {et~al.}(2004){Sparks}, {Donahue}, {Jord{\'a}n},
  {Ferrarese}, \& {C{\^o}t{\'e}}}]{Sparks04}
{Sparks}, W.~B., {Donahue}, M., {Jord{\'a}n}, A., {Ferrarese}, L., \&
  {C{\^o}t{\'e}}, P. 2004, \apj, 607, 294

\bibitem[{{Voit}(2005)}]{Voit05}
{Voit}, G.~M. 2005, Reviews of Modern Physics, 77, 207

\bibitem[{{Voit} \& {Donahue}(1997)}]{vd97}
{Voit}, G.~M. \& {Donahue}, M. 1997, \apj, 486, 242

\end{thebibliography}

\end{document}